# BOLD v4: A Centralized Bioinformatics Platform for DNA-based Biodiversity Data

Authors: Sujeevan Ratnasingham*, Catherine Wei, Dean Chan, Jireh Agda, Josh Agda, Liliana Ballesteros-Mejia, Hamza Ait Boutou, Zak Mohammad El Bastami, Eddie Ma, Ramya Manjunath, Dana Rea, Chris Ho, Angela Telfer, Jaclyn McKeowan, Miduna Rahulan, Claudia Steinke, Justin Dorsheimer, Megan Milton, Paul D. N. Hebert

*Corresponding author - email: sratnasi@uoguelph.ca

**Abstract**

BOLD, the Barcode of Life Data System, supports the acquisition, storage, validation, analysis, and publication of DNA barcodes, activities requiring the integration of molecular, morphological, and distributional data. Its pivotal role in curating the reference library of DNA barcodes, coupled with its data management and analysis capabilities, make it a central resource for biodiversity science. It enables rapid, accurate identification of specimens and also reveals patterns of genetic diversity and evolutionary relationships among taxa.

Launched in 2005, BOLD has become an increasingly powerful tool for advancing understanding of planetary biodiversity. It currently hosts 17 million specimen records and 14 million barcodes that provide coverage for more than a million species from every continent and ocean. The platform has the long-term goal of providing a consistent, accurate system for identifying all species of eukaryotes.

BOLD's integrated analytical tools, full data lifecycle support, and secure collaboration framework distinguish it from other biodiversity platforms. BOLD v4 brought enhanced data management and analysis capabilities as well as novel functionality for data dissemination and publication. Its next version will include features to strengthen its utility to the research community, governments, industry, and society-at-large.

# 1 Introduction

Efforts to document biodiversity changes rely heavily on access to robust data sources and tools. Essential reference data is increasingly sourced from the GenBank, GBIF, Catalogue of Life, and BOLD databases, each with a distinct mission and operational mode. GenBank is a repository of nucleotide sequences spanning all organisms and genes[1] while the Catalogue of Life provides an authoritative taxonomic hierarchy[2]. By contrast, GBIF facilitates access to species occurrence data, predominantly through the aggregation and mediation of specimen occurrence records from natural history collections and observational data[2][3]. BOLD interfaces with all three types of data and functions both as an aggregator and a repository. Like GenBank, it assembles sequence information from diverse taxa, but to focuses on standardized gene regions[4]. The significance of BOLD lies not only in its large DNA barcode reference library but also in its potential to transform the manner in which species are defined – through genic analysis[5]. BOLD's objective is to simplify the DNA barcoding process by making data more accessible and easily interpretable for a wider audience[4].

The BOLD identification engine further enhances the platform's utility[6]. It allows users to upload unknown DNA barcodes, which are then compared against BOLD's reference sequences to generate an identification. Users receive the closest species match, effectively automating the identification process.

## 1.1 A Central Database

BOLD is a centralized database for DNA barcode records which allows researchers and practitioners to access, store, and analyze DNA barcode data. This centralization brings several important benefits. First, it enables easy access and comparison of DNA barcode data across geographic regions, ecosystems, and taxonomic groups. Second, it facilitates collaboration among researchers and institutions by promoting data sharing.

Centralization has enabled standardized data formats, enhancing the accuracy and reliability of the DNA barcode records employed for species identification and classification. Consolidiating all DNA barcode data makes it one place makes it possible to track changes in biodiversity over time, detect and monitor invasive species, and identify key species and habitats that require protection[7]. As data volumes grow, the availability of comprehensive, reliable data will support practical conservation efforts and informed decision-making.

## 1.2 Data Standards

The application of DNA barcodes for specimen identification requires the construction of reference libraries which connect sequences from designated barcode regions (e.g., COI for the animal kingdom) to carefully identified voucher specimens. Reference libraries integrate molecular data with taxonomic information and geo-temporal data. This combination enables algorithmic species identification[8] while also providing high-quality occurrence data.

To qualify for recognition as a DNA barcode, the gene sequence must derive from one of the regions designated by the research community as the barcode marker for a particular domain of life: COI for animals; matK, rbcL, and trnH-psbA for plants[9]; and COI and ITS2 for fungi [10, 11] In addition, records must satisfy three criteria:; (i) < 1% ambiguous bases (= Phred score <20)[12]; (ii) primer sequences; and (iii) the country of origin. As well, the sequences must be > 75% of the length of the barcode region (i.e., 487 bp for COI) ,[9][10, 11]. While these criteria must be met to gain formal barcode status, records failing to meet any of these criteria can still be stored in BOLD.

Data submitters mainly drive the status of records within BOLD, but it supports them in maximizing data quality. Various tools help to identify anomalies or low-quality records. For example, submitted sequences of COI are translated automatically into amino acids and compared with a Hidden Markov Model (HMM) [13] of the COI protein to verify they derive from this gene. Sequences are also examined for stop codons to aid the detection of sequencing errors, pseudogenes, or chimeras. Contamination checks are performed by comparing barcode genes (COI, matK and rbcL, ITS2) against known laboratory contaminants. For animals, this includes several vertebrates (human, mouse, cow, pig) and known endosymbionts, among others. The record is flagged if a close match to a contaminant is found. BOLD also captures the event history of records to ensure that erroneous modifications can be identified and reversed.

**1.3 DNA-Based Species Identification**
BOLD's identification engine is a key feature as it allows users to identify unknown specimens based on their barcode sequence(s). This functionality employs sequence similarity thresholds to generate identifications. Although more rigorous methods are available [14, 15], BOLD employs similarity methods because they are more computationally efficient, and their accuracy is adequate for most use cases[16].

BOLD employs two methods to deliver identifications. For animals, a two-stage process is used to assign an identification. Query sequences are first translated and aligned using a profile Hidden Markov Model[17] of the COI protein. Identifications are then generated by calculating distances between each query sequence and all those in the reference library. The 100 records in the reference library with the closest match are returned along with an assignment to a taxon at a given rank based on similarity thresholds. A single-stage process, based on the BLAST+[18] algorithm, is used for all other barcode markers.

The reference sequence library for each gene region is updated weekly, incorporating new submissions to BOLD and NCBI. Most records on BOLD are public but some are private. Specific data elements are redacted from private records to safeguard the rights of the data owners, ensuring they can publish at a reasonable pace. However, these private records are immediately included within the identification engine so all users benefit from the most recent data and to aid the detection of erroneous records.

Five reference sequence libraries are available (**Table 1**) to support different needs. Four are updated weekly while the Historical libraries represent an annual snapshot.

| Barcode Library | Conditions For Inclusion | Target Use Cases |
| --- | --- | --- |
| Species Level | Records meet the barcode standard. Source specimens have a species designation. | This library delivers species-level matches. Some matches are to private records. |
| Public | Records meeting the barcode standard which are public | This library should be used when access to the records underlying an identification is required (e.g., regulatory or forensic uses). |
| Species Level Full Length | Records with a species-level taxonomic assignment and a full-length barcode (> 640 bp for COI) | This is the highest quality library on BOLD as it only includes records with species designations and full-length barcode sequences. This library is very effective for the identification of short barcode sequences. |
| All | All records meeting the barcode standard. | This library is the most comprehensive, but many records possess limited taxonomic resolution (e.g., identified to a family). |
| Historical | A snapshot of the records in the library at the end of each year | These libraries focus on reproducibility and support re-evaluation of prior studies. |

**Table 1.** Reference libraries on BOLD and their target use cases. The BOLD Identification Engine is accessible from three points on the platform. The commonest access point is from the Identification Page (http://www.boldsystems.org/index.php/IDS_OpenIdEngine) which is open to any user. A more scalable option is available through the Workbench. Its use requires registration, but it supports the batch identification of hundreds of sequences. The final option is via the Application Programming Interface (API) which allows users to programmatically engage a service endpoint to request identifications, enabling them to integrate identification services into a pipeline or custom application.

### 1.4 Workbench Supporting Data Collection and Publication

BOLD manages DNA barcode data from the earliest point in its lifecycle. Its primary function is to provide secure, centralized storage of DNA barcode sequences and their associated metadata. It employs strict data standards to enable easy retrieval and comparison of records from various sources. BOLD also provides analytical and quality control tools to help researchers explore and analyze the data, streamlining the overall management process.

BOLD is designed to facilitate collaboration among researchers working with DNA barcodes. It meets this goal by providing a simple path to create projects and assigning roles and permissions to team members, enabling seamless data sharing. Furthermore, it supports data visualization and export, helping researchers communicate their results to a broader audience.

In addition to fostering collaboration, BOLD supports the publication of DNA barcode records. Researchers can publish records within BOLD, enabling other researchers to integrate them into broader analyses. As well, it eases the submission of data to GenBank by handling all aspects of the submission process, including data reformatting and deposition. These functionalities make BOLD an invaluable tool for researchers, as it enables the sharing and dissemination of data in a standardized format.

**1.5 Registry for Dark Taxa**

The Barcode Index Number (BIN) system[5] is a key feature of BOLD. Its development was propelled by a need to address dark taxa[19], the millions of species lacking a scientific name. By providing a standardized method for clustering sequences into persistent operational taxonomic units (OTUs), the BIN system brings objectivity and reproducibility to species delineation and identification, bridging the gap between traditional and molecular taxonomy.

The BIN system employs an algorithm that combines single linkage and Markov clustering to group DNA barcode sequences into OTUs based on their similarity. Clustering is performed primarily on COI sequences, most derived from animals. The system utilizes the Refined Single Linkage (RESL) algorithm[5], which considers both cohesion within a group of sequences and their stability to enable a highly reliable method for species delineation.

The BIN registry is updated monthly in response to new data and modifications to existing records. When a new BIN is recognized, it is assigned a unique alphanumeric identifier and a corresponding webpage is established on BOLD (**Figure 1**). This page assembles all records for the BIN, and aggregates information on their taxonomic assignments, images, and collection points.

The effectiveness of the BIN system as a proxy for species has been demonstrated in diverse taxonomic groups[20–23]. It employs Uniform Resource Identifiers (URIs) and web service functionality to support integration with other databases such as GenBank and GBIF. Deeply embedded in BOLD, BINs are readily accessed via consoles and analytical tools.

Since its launch, the BIN system has gained broad adoption, being cited in more than 2,000 publications. It plays an increasingly important role in accelerating taxonomic research and species descriptions[24, 25]. It also makes it possible to evaluate patterns of biodiversity in groups that have received little taxonomic attention. With continued application and further development, the BIN system promises to fast track the registration of all animal species.

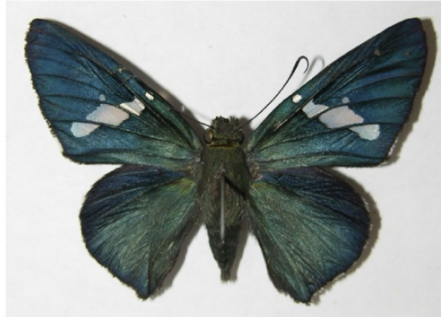
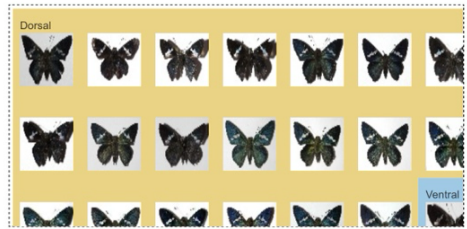
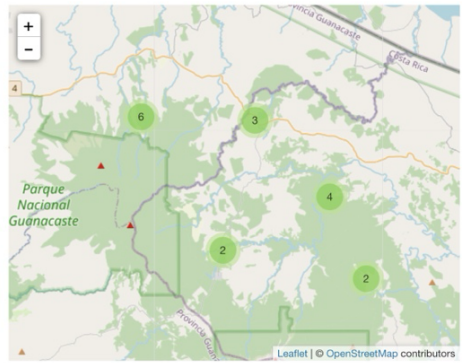
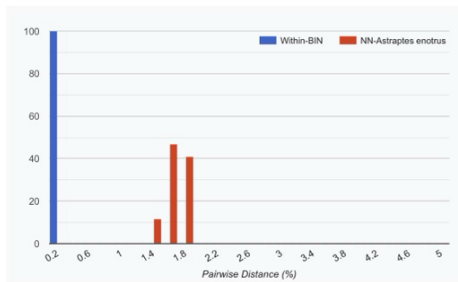

**Figure 1.** BIN page for BOLD:AAC1842 which represents an interim species, *Astraptes janeriaDHJ02*.

## 2 Data Model

BOLD employs a unique data model (**Figure 2**), one combining DNA sequence data with fields traditionally captured by museums (image and site/date of specimen collection). The initiation of each record requires two data elements (a unique Sample ID (e.g., BIOUG30290-H09) and a high-level taxonomic assignment (e.g., phylum). Once the Sample ID has entered BOLD, further data elements can be added as they become available. Moreover, the extension of existing records with new data (e.g., sequences for more genes) does not impact populated fields. Flexible curatorial workflows are supported allowing access to distinct data elements can be managed independently. For example, permissions to modify the identification of specimens can be provided to taxonomic specialists, while being withheld from molecular specialists who would, on the other hand, be granted access to modify sequence data.

**Specimen**
- * Sample ID
- Field ID
- Museum ID
- Recordset code array

**Attributes**
- Sex
- Life stage
- Extra info
- Tissue type
- Reproduction
- Storing institution
- Voucher type
- Sampling protocol
- Associated specimens
- Associated taxa
- Funding source
- Specimen linkout

**Taxonomy**
- Kingdom
- * Phylum
- Class
- Order
- Family
- Tribe
- Genus
- Species
- Subspecies
- References
- Identified by
- Identification method
- Identifier email
- Notes

**Collection Location**
- Country
- Country code
- Province
- Region
- Sector
- Site
- GPS coordinates
- Elevation/Depth
- Location accuracy
- Collectors
- Collection date
- Collection time
- Collection accuracy
- Collection code
- Site code
- Collection event ID
- Notes

**Images**
- Orientation
- Attribution
- Licensing

**Sequence**
- Process ID
- Genbank accession number
- Sequence upload date
- ProcessID minted date

**Marker Sequence**
- Gene
- Sequence
- Marker code

**Primers**
- Forward primer sequence
- Forward primer code
- Reverse primer sequence
- Reverse primer code

**Traces**
- Quality scores

**BINs**
- Unique identifier
- Nearest Neighbour
- Sequence variation

**Figure 2.** The BOLD data model with specimen-related data elements in green and sequence-related ones in orange. * Fields required to instantiate a record.

### 2.1 Specimen Data

BOLD provides a page that consolidates all data elements associated with each specimen (**Figure 3**). This Specimen Page serves as an up-to-date representation of the record, instantly reflecting any modifications to the underlying data elements. To speed curation, all components of each record can be edited directly from the Specimen Page.

Modifications to records are tracked through versioning, ensuring a comprehensive audit trail. The system maintains a history of changes in two ways: (i) An activity log tracks change events and the user(s) responsible for them, offering a comprehensive chain of modifications; (ii) The Delta View captures and displays the history of each record, highlighting changes (or deltas) between weeks, enabling efficient monitoring and comparison of alterations to each record over time.

Directly modifying fields is just one way to manipulate data on BOLD; each specimen record can also be annotated with tags and comments. For instance, if a user believes the identification of a specimen is incorrect, a tag can be added, such as "Suspicious Identification," as an alternative to removing the present taxonomic assignment. This function allows for further contextualization and collaboration among users.

By offering direct editing capabilities, comprehensive change tracking, and annotation, the Specimen Page ensures efficient data curation, enhancing the overall usability and reliability of the system.

Specimen images, although optional, have become a critical component of records on BOLD as they play a crucial role in quality assurance and taxon profile construction. They also enable the inspection of morphological characters, enhancing the credibility of taxonomic assignments.

Guidelines are provided to ensure the comparability of images. Employing strict image orientation and standard aspect ratios enables the simultaneous comparison of thousands of images by constructing image arrays. Images are often uploaded to BOLD before sequence analysis, enabling their use in sequence validation.

Image licensing is determined by the image owner and can be modified. The International Creative Commons - Attribution – Share Alike licence is advocated as it balances protection and accessibility. It is used as the default licence when none is provided.

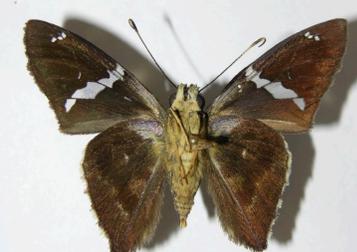

**Figure 3.** Specimen Page. This Specimen Page for an individual of *Atraspes tucuti* (Hesperiidae) shows the seven standard elements: 1. Identifiers and attributes; 2. Photograph(s); 3. Image annotations as tags and comments; 4. Collection site; 5. Specimen details, including links for editing specimen information and to access the Delta View. 6. Activity report. 7. Comments. All elements on each Specimen Page can be modified via the Edit Specimen function available below the specimen details.

## 2.2 Sequence Data

The Sequence Data page provides a consolidated view of all fields of a sequence record together with supporting information (**Figure 4**). Accordingly, it displays the code(s) for the targeted gene region(s), the sequences, the PCR primers used to amplify them, as well as the trace files. When multiple markers are submitted for the same specimen, the data for each marker is displayed separately. Similar to the Specimen Page, users can modify every element of the sequence data directly from the Sequence Page. Modifications are tracked through an Activity Log and Deltas.

Sequences frequently feature associated electropherograms (traces) and, if available, quality scores (e.g., Phred). With the broad adoption of second and third-generation sequencing, the support for the reads in FastQ format[26] underlying their consensus sequences will soon be added. While these files are necessary required for records to attain barcode status, they provide valuable information about sequence quality. BOLD records failed traces for record keeping, while other results are placed in three quality classes (Low, Medium, High). Prior to uploading traces, specimen records must be established, and the primers used to generate the sequences must be registered in the BOLD Primer Database.

The Primer Database is a registry of all primers employed to amplify a barcode region (e.g., COI, matK, rbcL, ITS2). It is mandatory to register primers before assembling the sequence submission package. When an uploaded primer sequence matches an existing sequence with a different name, BOLD reuses the original sequence and provides its registered code for association with sequence records.

Users can upload and update sequence records at any time. Once on BOLD, sequences are instantly viewable, downloadable, and ready for analysis.

**Figure 4.** Sequence Page for a representative of *Atraspes tucuti* (Hesperiidae) with the following information: 1. Specimen details; 2. Links to Activity Report on the sequence and a Delta View. 3. Marker(s) summary: when there are additional markers, their information is displayed independently; 4. Marker(s) information details, including the nucleotide and amino acid sequence and its metadata. As with the summary, when there is more than one marker, their information is displayed independently. 5. Trace file(s) are displayed and available for download.

## 3  Data Management

BOLD employs a flexible data management and access model to support a high volume of data and diverse workflows. The system utilizes a hierarchical structure to support the organization of records (**Table 2**). After login, users are directed to the Main Console, a command centre that facilitates data monitoring and management. From it, users can create new projects or access existing ones. Since projects serve as the primary management structure on BOLD, all new records require a designated home project. Each project is equipped with an access control list (ACL) under the control of the Project Manager (the user who created the project). The ACL supports multiple access levels (**Table 3**), ranging from the most restrictive (read-only) to the most permissive (modify any data field).

| Structure | Definition | Use Cases |
| --- | --- | --- |
| Project | As the fundamental management unit, it holds records and supports access control. Each project has a manager who controls access to it. All new records require a project for deposition and maintain this affiliation through a shared code. | Data generation and validation. Collaboration with a small team or within a single laboratory. |
| Folder/ Container | Folders have similar interfaces to projects but contain projects instead of records. Containers have separate access control allowing data to be shared with users operating across a set of projects. | Data lifecycle management. Collaboration across small networks. Reporting and auditing. |
| Campaign | Provides an organizational structure that groups folders and projects based on theme or initiative. This structure does not directly support access control but provides a way to affiliate projects for reporting or analysis. | Reporting. Communications. |
| Dataset | Provides a container for records deriving from multiple projects with a taxonomic, geographic, or thematic focus. Datasets are typically used to support publications due to their ability to mint DOIs. | Publication. Data curation. |

**Table 2.** Data management structures supported by BOLD.

| Access | Definition | Appropriate Use |
| --- | --- | --- |
| Analyze | Allows users to execute various analyses on the sequences in a project. | The option should be made available to all users on a project. |
| View & Download | Allows users to view sequences and download them. | This option should be made available to users who require the capacity to export data for analysis using third-party tools or for sharing outside BOLD. |
| Edit Sequences | Allows users to edit, overwrite, and delete sequences. | This option should be made available to those generating sequences as it allows them to upload and update sequences. It also allows data curators to repair or excise erroneous sequences. |
| Edit Specimens | Allows users to modify any data element associated with the specimen record including taxonomy and geography. | This option should be made available to those assembling specimen records and to those with taxonomic expertise. |
| Add to Dataset | Allows users to add new records into a dataset under their control, giving them the ability to share or publish the records. | This option should only be made availability to users who require the capacity to add records into datasets for publication. |

**Table 3.** Five options which control access to varied domains of the records in a project. Each option can be activated independently except Add to Dataset as it also includes the Analyze and View & Download options.

While individual projects often meet the needs of a single researcher or small teams, large-scale projects and research networks necessitate additional structures to support complex data ownership, lifecycles, and reporting. Creating multiple projects and organizing them using the Folder or Campaign structure on BOLD is advisable in such cases. The folder structure aids project organization and introduces an additional ACL independent of its contained projects. This functionality is particularly useful when periodic review is required by individuals who are not directly involved in routine management of the data.

Research networks often involve dozens of researchers, and effective data management necessitates multiple projects and folders. In such cases, reporting becomes challenging as the project count can exceed 100. The campaign structure was explicitly designed to address this situation. Although lacking an ACL, the campaign structure includes an Application Programming Interface (API) that provides data feeds to national or network databases and dashboards external to BOLD.

The final structure (Dataset) is available to assemble records from diverse projects, often in support of a publication. Although Datasets fall outside the hierarchy of the other data management structures, they employ an ACL. Unlike other structures, they can pool records from multiple projects and folders to provide maximum flexibility and integration.

**3.1 Projects**

Projects are the core management unit within BOLD as they provide a secure means of data management. This organizational level streamlines workflows by providing an efficient environment for loading, storing, and tracking records. Each project is identified by a unique code of three to five letters that also forms the prefix for the ProcessID affiliated with every record in it.

The Project Management console (**Figure 5**) grants real-time access to key components of the project. It enables immediate monitoring of newly added records, verifies the completeness of data, and swiftly detects quality issues. Importantly, this console does not simply identify error identification; it allows their correction. Upon identifying data quality concerns, it generates clickable links that direct users to problematic records, expediting resolution of errors.

Access to the Project Console and the data within each project are strictly controlled via an ACL that serves a dual function. It upholds data integrity by restricting access to authorized users and lists them together with their respective access levels. Project managers possess the capability to revise these access rights, allowing them to alter the ACL to meet new requirements for the project whether this involves revoking access rights to some existing users or conferring permissions to new ones.

An essential feature of each project is the production of an Activity Report. This report provides a detailed log of all data uploads and modifications made to the records within it. Offering a transparent audit trail, it guarantees traceability. The Project Console includes a range of analytical tools accessible through the main menu to facilitate data exploration and validation.

BOLD provides a drill-down function termed the Record Console **(Figure 6)** for detailed data inspection. This feature allows users to view and interact with individual records within a project. Records can be accessed via links leading to their Specimen or Sequence Pages. Links are also available to BIN pages when a record has been assigned to one. Additionally, each record is annotated with error and compliance flags which assist in identifying issues such as contamination, the presence of stop codons (for protein-coding sequences), misidentification, and non-compliance with barcode standards.

The Record Console uniquely allows users to select a subset of records within a project for download, for analysis using the integrated tools, or for addition to a dataset.

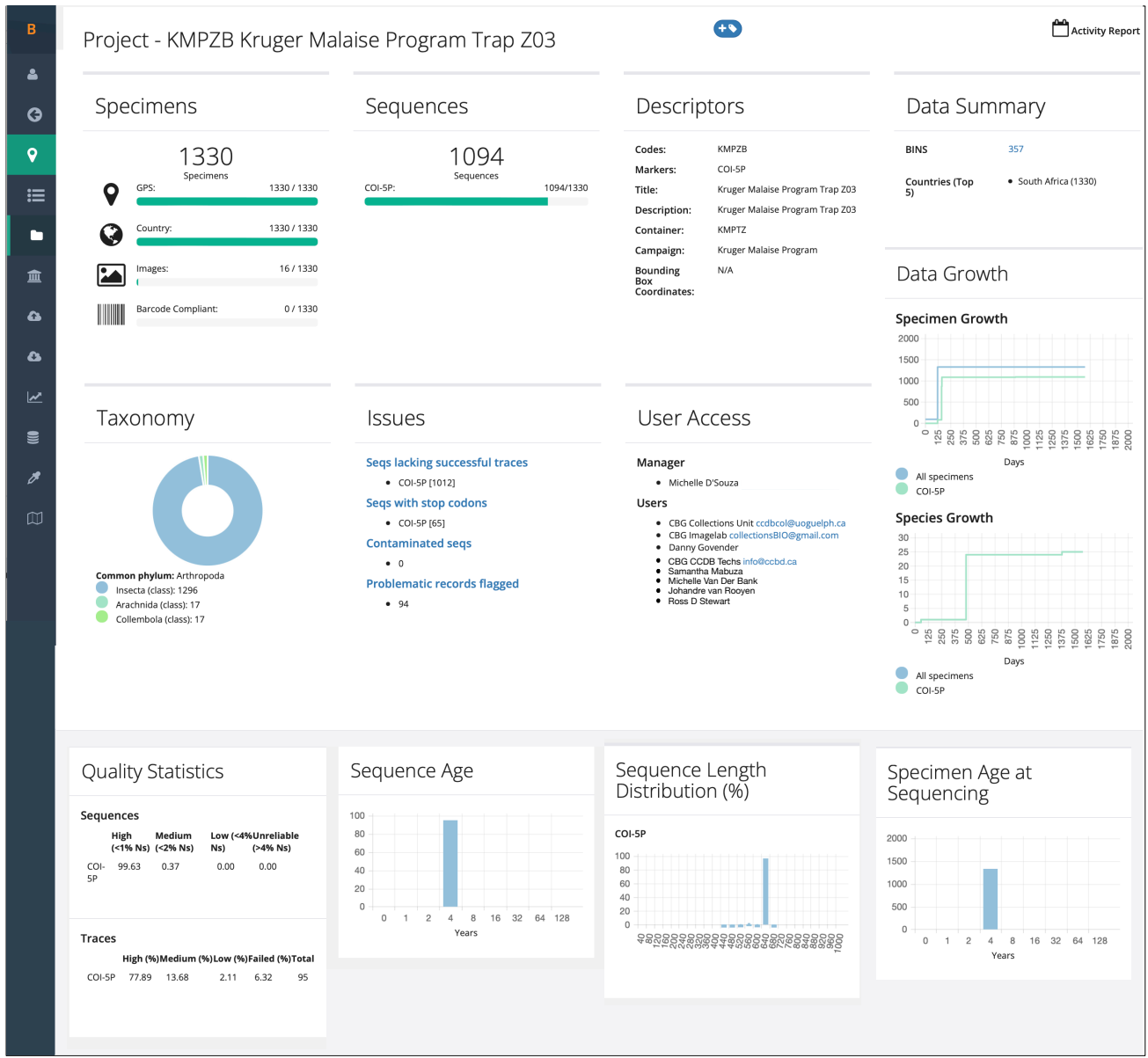

**Figure 5. Project Console.** The Project Console provides an overview of key metrics including the number of specimens, sequences, and BINs associated with each project. It also displays important statistics such as the success in sequence recovery, the BIN discordance rate, and a histogram of sequence lengths. Finally, it highlights issues associated with records in the project.

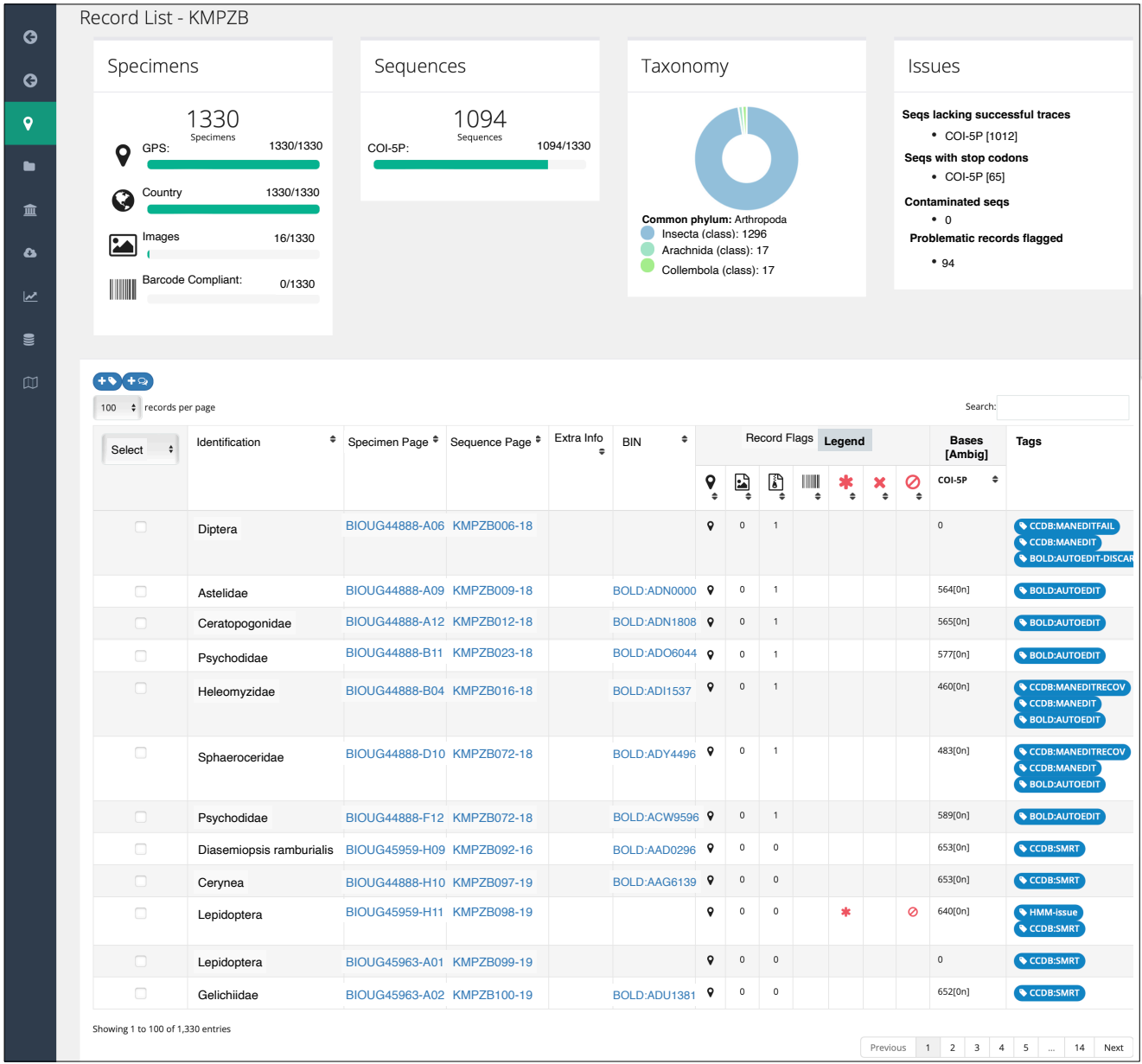

**Figure 6. Record Console.** The Record Console provides general statistics about the composition and completeness of the records in a project. In addition, a summary of detected issues is highlighted. A data table displays individual records and their features (presence of GPS coordinates, images, trace files, if they are barcode compliant, stop codon, contamination, flagged). It also displays the tags (if any) associated with each record. Records and associated BINs are accessed by clicking on their identifiers.

## 3.2 Datasets

The Dataset feature on BOLD significantly enhances data organization and accessibility. It shares some functionalities with those available in Projects but several key differences make it a valuable supplement to BOLD's data management system.

In contrast to Projects, which function as the primary record containers, Datasets operate as independent collections. They allow users to categorize and assemble records based on multiple criteria or to support specific research goals. Although any user can create a Dataset with published records, the inclusion of private records requires permission from the Manager of the Project containing those records.

Once a dataset is created, its manager can share it with other users. Dataset access is governed by an ACL, akin to that used for projects. The dataset ACL manages who can access and interact with the Dataset, reinforcing the platform's dedication to data integrity and secure collaboration.

The dataset feature incorporates functionalities available to conventional projects, including analysis tools and drill-down capability. Each dataset is accessed via the Dataset Console, similar to the Project Console but with additional functionality. One key feature is the ability to generate Digital Object Identifiers (DOIs) to facilitate data packaging for publications. The resultant DOIs can be directly cited in the manuscript, offering a straightforward, robust method for linking to the data. Each DOI provides an immediate link to the dataset, allowing reviewers to view and download the relevant data. The integration of DOIs within datasets enhances data discoverability and traceability (Mannheimer 2016), making it an invaluable tool for scientific publications and data sharing.

## 4 Data Lifecycle

BOLD incorporates a life cycle approach to data management [14]. It facilitates the integration of specimen information and molecular data, such as sequences, primers, and traces along with other elements like images and maps in a centralized repository. This structured, unified approach helps to eliminate data silos, enhances data sharing and interoperability, and increases resilience [15] (Gilman et al., 2009). Notably, each phase of the record life cycle can be performed at any time, supporting dynamic data transformations based on updates and evolving requirements. Detailed discussions about the data life cycle (**Figure 7**) follow this section.

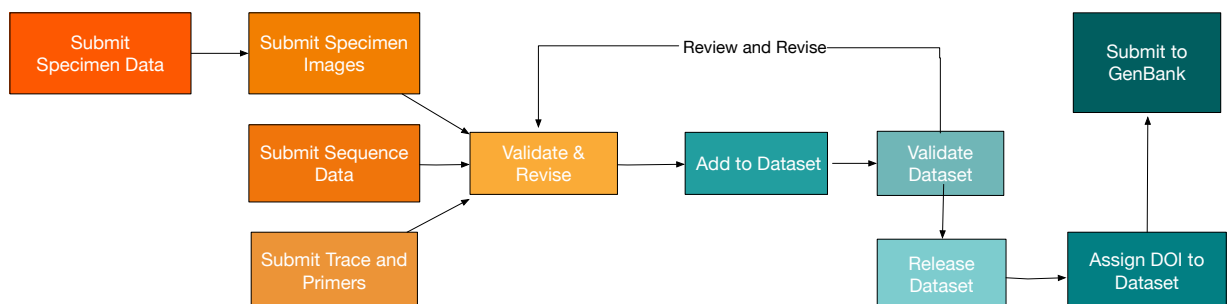

**Figure 7. Data Lifecycle** The data lifecycle on BOLD starts with specimen data and ends with the publication of records and submission to public repositories.

## 4.1 Specimen Submission

The assembly of a complete record involves multiple stages but commences with specimen submission. To establish a record, certain minimum data elements are required, including (i) a unique Sample ID; (ii) Field ID and/or Museum ID, the identifiers assigned to a specimen in the field or upon institutional accessioning; (iii) the name of the storing institution; (iv) the specimen's phylum; and (v) the country of collection. Additional elements can be incorporated/updated throughout the record's lifespan.

BOLD accepts submissions of specimen data through an online form for single or small batches of records or via a spreadsheet (http://www.boldsystems.org/index.php/Resources/Templates) for larger volumes. Although registered users can upload information to create BOLD records, a project must already exist or be created to house the data. As soon as records are generated, they are instantly viewable and editable by those with appropriate access to the project.

BOLD's ability to extend and update records with taxonomic, geographic, and other specimen data can be employed at any stage after record creation. The submission workflow allows extensions and updates in three groups: specimen descriptors, taxonomic assignments, and collection information. Updates submitted to BOLD replace the existing information with new data, but a historical record persists. BOLD tracks the details of each update (who, what, when) to ensure a comprehensive audit trail.

Most submissions to BOLD proceed smoothly, but inaccurate or inconsistent data will prevent the upload process. BOLD conducts checks on all submissions to ensure the data is of high quality and fit for purpose. Data issues fall into three categories: consistency, completeness, and compliance (**Table 4**) and have prescribed remedies.

| Categories | Checks | Resolutions |
|---|---|---|
| Consistency | GPS coordinates in the correct format. | Format errors are corrected where possible and the submission proceeds. If a correction is not possible, the submitter is contacted. |
| | Geographic names consistent and matching registry | |
| | Date in the correct format. | |
| | Specimen depository matching one registered on BOLD. | New depositories are added once registration information is acquired and the submission proceeds. If information is unavailable, the submission is paused until the submitter provides it. |
| | Taxonomic nomenclature must match the BOLD backbone taxonomy. | New scientific names are verified by searching the literature. If found, they are added to BOLD, and the submission proceeds. If not, the submitter must provide a reference or withdraw the name. |
| | The Identifier (individual providing taxonomic assignment) must be registered on BOLD with an affiliation. | The Identifier's affiliation is ascertained and the submission proceeds. If this is not possible, the submission is paused until the submitter provides this information. |
| **Completeness** | GPS coordinates provided. | Submission continues, but the record is annotated to indicate that key information is missing. |
| | Collection dates provided. | |
| | Taxonomic names provided with no gaps in the hierarchy. | Gaps are filled where possible based on BOLD's backbone taxonomy. Where a conflict is encountered, the submission is paused and the submitter is notified. |
| | Specimen identifiers (SampleID, FieldID, MuseumID) are unique and of sufficient length to maintain uniqueness. | Submissions failing to meet this criterion proceed only once addressed. |
| **Compliance** | Unique primary sample identifier (Sample ID). | Submissions are paused until these compliance criteria are met. Exceptions are made in some cases (e.g., obfuscation of GPS for endangered species). |
| | Country specified. | |
| | Specimen depository specified. | |
| | Collection coordinates at minimum precision (two decimal) | |

**Table 4.** A list of common issues encountered during the submission of specimen data and their resolution.

Submissions introducing terms new to BOLD require communication with BOLD staff. For example, adding new taxonomic names may lead to a request for supporting information. While such cases pause the process, they expand BOLD's controlled vocabularies or libraries, ensuring future submissions proceed smoothly. However, most triggers for investigation result from typographic or data entry errors which must be corrected before submission proceeds. These checks are critical to maintaining data quality.

### 4.2 Image Submission

Images are a very important part of BOLD records because of their value in validating taxonomic assignments[27, 28] and as a source for trait data[29]. To maximize their utility, images should be uploaded early in the record assembly process. BOLD makes use of the images in two ways: (i) quality assurance, supporting the verification of taxon assignments (ii) constructing taxon profiles, with BOLD displaying a subset of images in the Taxonomy Browser and on BIN pages. Plans are underway to leverage these images to facilitate taxonomic classification through AI methods[30].

Images may be submitted individually or in batches. Batch submissions require a zip package containing an Excel File with the necessary information (**Table 5**) and a collection of jpeg images. Batch submissions can be performed via the Main Console or the Project Console. BOLD accepts high-resolution images of up to 20 megapixels with a maximum of 200 images per package. Larger submissions are permissible but require prior contact with BOLD's support team (accessible via support@boldsystems.org).

| Data Types | Field Name | Definition |
|---|---|---|
| Identifiers | Image name * | Name of the file with the .jpg extension. It should be identical to the image file name. |
| | Sample ID | Identifier of the specimen record previously created in BOLD. |
| | Process ID | Identifier of the record sequence in BOLD. |
| Attributes | Original specimen * | 'Yes' if the image shows the actual specimen of the record. 'No' if the photograph is a representative specimen of the same species, a photograph of the collection site, etc. |
| | Orientation * | Position of the specimen in the image. For example: dorsal, ventral, or lateral. |
| | Caption | Free text description of the image with a limit of 400 characters. |
| | Measurement | Any relevant measurement in metric units. |
| | Measurement Type | Feature that was measured. |

| License | License Holder | The primary individual who holds the licence. |
|---|---|---|
| | License * | BOLD supports eight (8) types of image licence.<br>Copyright<br>No Rights Reserved<br>CreativeCommons Attribution<br>CreativeCommons Attribution Share-Alike<br>CreativeCommons Attribution No Derivatives<br>CreativeCommons Attribution Non-Commercial<br>CreativeCommons Attribution Non-Commercial Share-Alike<br>CreativeCommons Attribution Non-Commercial No derivatives |
| | License Year | The year of the licence declaration. Limit of four characters. |
| | License Institution * | Institutional affiliation of the primary license holder. If the individual holder is unresponsive/unreachable, decisions regarding the use of material fall to the institution. |
| | License Contact | Contact information for the license holder. Email address, mailing address, phone number or all the above. Limit of 128 characters. |
| Attribution | Photographer | The individual or team responsible for photographing the images. Limit of 128 characters. |

**Table 5.** Metadata associated with images on BOLD (* Denotes mandatory fields).

Each image requires a license with the default being the Creative Commons - Attribution – Share Alike which deftly balances protection and accessibility. In the absence of a provided license, the default option is applied automatically.

BOLD provides a standard guide for photographing specimens. Although dorsal, lateral, and ventral views are ideal for animals, candid or unoriented specimen images are of considerable value. BOLD supports the inclusion of captions or other descriptors for each image. Photographs where diagnostic features are measured or displayed can often prove beneficial for identification. For plants, full-sheet photographs or scans of herbarium sheets with close-ups of diagnostic features are ideal (e.g., **Figure 8).**

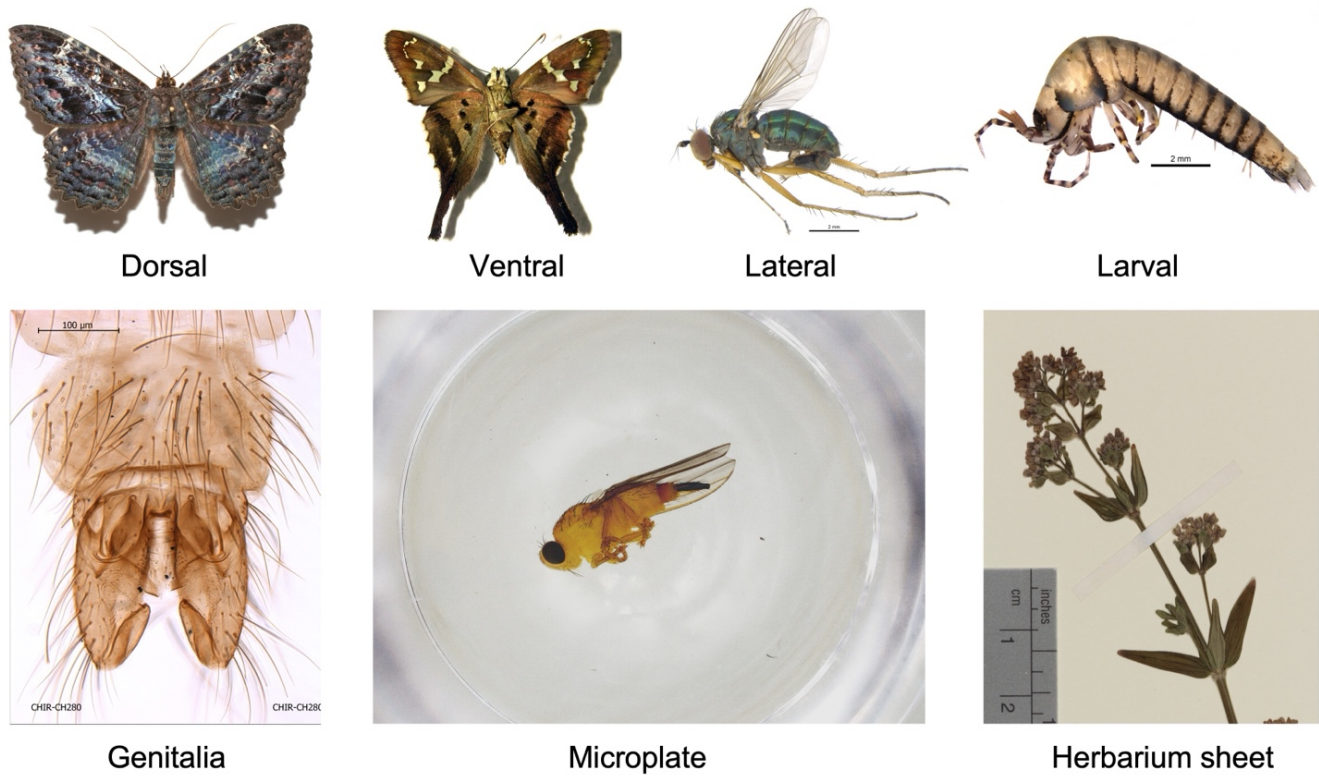

**Figure 8**. Example specimen images and commonly utilized orientations. Dorsal, Ventral, Lateral, and Larval , Genitalia, Microplate, and Herbarium sheet  orientations are depicted.

Up to ten images can be uploaded for each specimen, allowing users to associate their records with images of live or mounted organisms, distinct features, hosts, or habitats.

Submission templates are available at http://www.boldsystems.org/index.php/Resources/Templates.

**4.3  Sequence Submission**
The addition of sequence data is the final step in the assembly of a record. In order for sequences to be uploaded, their source specimen(s) must already be registered. The sequence region or marker should also be registered in BOLD's marker database. New markers can be registered by contacting BOLD's support team.

Upon submission, sequences are immediately associated with records. In addition, BOLD automatically translates protein-coding sequences, flagging those producing stop codons. It also compares uploaded sequences for barcode regions (e.g., COI), against a contaminant library, such as human, mouse, cow, pig, etc. If the system detects a close match, it flags the records for review.

Sequence data can be uploaded via two access points: (i) The first and most frequently used is via the Project Console. (ii) The second entry point is via the Main Console, where a sequence upload option appears in the dashboard. Uploads require specific metadata (**Table 6**), including the marker being uploaded and the laboratory which generated the sequences. BOLD accepts up to 2,000 sequences in a single upload. BOLD also allows for sequence editing, deletion, and replacement directly from the sequence page. To modify a sequence, navigate to the Sequence Page for the target sequence and click the Edit Sequence button.

| Field | Definition |
| --- | --- |
| **Identifier** | Each sequence requires a unique identifier. Two options are available: SampleID and ProcessID. |
| **Markers** | The gene region from which the sequence originates. The name should be registered in BOLD's marker database (255 markers currently registered). Sequences from only one marker can be uploaded at a time. If the marker is not provided, the system will use the default value – the primary marker associated with the project. |
| **Run site *** | Name of the facility responsible for generating the sequence(s). If the facility is new to BOLD, it must be registered. Sequences from only one facility can be uploaded at a time. |
| **Sequence *** | One or more sequences in FASTA format. Ideally, sequences should be aligned, but this is not required.<br><br>The FASTA defline must conform to a strict format. It should start with '>', followed by the ProcessID or SampleID. Other data may be included in the defline but must be preceded by a pipe ('\|') to mark the termination point of the ProcessID. For example:<br>*>ProcessID\|Other info*<br>ATGGCCTTTAATCG… |

**Table 6. Sequence information.** Data associated with sequence upload. Mandatory fields are denoted with an asterisk (*).

### 4.4 Trace and Primer Submission
Trace files and primers help to ensure the verification and reproducibility of sequences submitted to BOLD. In addition, they fulfil a compliance requirement set by the BOLD data standard. Prior to the upload of trace files or primers, specimen records must be established.

Given their interrelated nature, trace files and primers associated with sequence records are uploaded in tandem. To register new primers, users need to complete a form with the information outlined in **Table 7**, which is accessible via the Data Uploads menu in the Main Console.

| Field Name | Definition |
| --- | --- |
| **Primer code** | Code for the primer. Between 3 and 12 characters. |
| **Marker** | Gene region that the primer set amplifies. It should be one of the registered markers in the BOLD database. |
| **Cocktail** | Standard combination of multiple primers used in one PCR reaction. Two options: 'Yes' or 'No'. |
| **Primer Sequence (5' to 3')** | DNA sequence for the primer in the 5' to 3' direction. |
| **Direction** | Two choices: 'forward' or 'reverse'. |
| **Reference/Citation** | Publication associated with the primer if one is available. |

**Table 7.** Primer information. This table displays data required for primer registration.

It is best to confirm that the primers linked to the trace files are registered in BOLD's Primer database prior to uploading. There are two main methods to submit this data to BOLD: (i) Single Record Trace File Upload: Available from the Sequence Page, this option allows you to upload trace files for a single record directly from its Sequence Page. Ideal for a small batch of trace files, this feature accommodates a maximum of four electropherograms and four optional Phred/score files per locus for each record. The association between traces and the records is immediate upon submission. (ii) Batch Trace File Upload: This method requires grouping trace files into a folder, coupled with associated metadata (**Table 8**) in a spreadsheet form, and creating a single zip file for upload. The batch method permits up to 10 trace files per record, with a total limit of 400 files per package. All files included in the upload must be listed in the associated spreadsheet form.

| Field Name | Definition |
| --- | --- |
| **Trace File** | Complete name, including extension ('ab1' or 'scf') identical to the file name in the folder. |
| **Score File** | Complete name, including extension (.phd.1) identical to the file name in the folder (if available). |
| **PCR Primers Fwd/Rev** | Primer codes for the forward and reverse primers. Both fields must be present. Primer codes are case sensitive. |
| **Sequencing primer direction** | Forward or Reverse. |
| **Sequencing Primer** | Primer used for sequencing. |

| **Process ID** | Identifier of the sequence in BOLD. |
|---|---|
| **Marker** | Gene region amplified by the primer set. It should be one of the registered markers in BOLD. For a multi-marker project, this field must be completed, If not, the traces will be uploaded to the primary marker. |
| **Sequencing Lab** *(only for form-based trace for single record)* | Name of the facility generating the sequence(s). If the institution is not listed, it must be registered. |

**Table 8.** Trace file information. Data required for trace file submission.

### 4.5 Record Validation

Validation is a critical phase in the life cycle of every record on BOLD. This stage employs a blend of automated and manual checks to ensure the completeness and integrity of each record.

There are two stages to validation. Initial validation occurs during data upload, but it is followed by a second stage during reconciliation with other records. During the first stage, BOLD largely automates the process, checking each record for completeness and comparing data against controlled vocabularies. This system swiftly flags discrepancies, allowing submitters to rectify errors. For instance, country names not recognized by the database are checked for spelling errors. BOLD performs similar automated checks on specimen data during upload, repairing or removing erroneous data elements after their recognition. However, automated detection can be difficult due because of the complexity of certain data fields such as species names or detailed descriptors like the Exact Site. As a result, responsibility for comprehensive quality inspection rests with the data submitters and project managers.

Errors identified in specimen data usually involve typographic or transliteration mistakes which can be easily corrected and are often addressed automatically. However, sequence data errors require a more in-depth investigation before resolution. For example, protein-coding sequences that yield stop codons upon translation could indicate sequencing errors or the presence of pseudogenes, necessitating the inspection of the sequence and its comparison with sequences from related organisms.

The second validation stage commences once specimen and sequence data have been uploaded to the record. This stage involves an evaluation of the record's validity by comparing it with existing records, either from the same project or from published data. BOLD performs some of these checks in an automated manner, such as scanning new sequences against a contaminant library. However, much of this validation work is carried out by users through data analysis. To assist with user-driven record validation, BOLD provides a suite of integrated analytical tools. Frequently used tools include the Taxon ID tree, Distribution Map, and Batch ID Engine.

The Taxon ID tree tool generates a Neighbor Joining tree[31, 32], encompassing all project records. The tree output, labelled with record taxonomy, enables easy comparison between tree topology and record taxonomy. Selecting the Matching Images option enhances this tool's utility as it generates an image array that is in the same order as the records in the tree (**Figure 9**). This feature provides a morphological context to tree placement, offering an immediate indication of misplaced nodes. It also organizes images based on their sequence similarity, making instances of contamination or misidentification readily apparent due to their contrast with neighbouring images.

The Distribution Map tool projects specimen coordinates onto a map, facilitating the swift identification of errors in coordinate values. This tool also supports direct correction of errors from points on the map, streamlining the removal or correction of erroneous coordinate data.

Lastly, the Batch ID Engine complements the function of the Taxon ID tree. It compares sequences in a project against all data on BOLD, detecting discrepancies based on the assumption that established records on BOLD have accurate taxonomy.

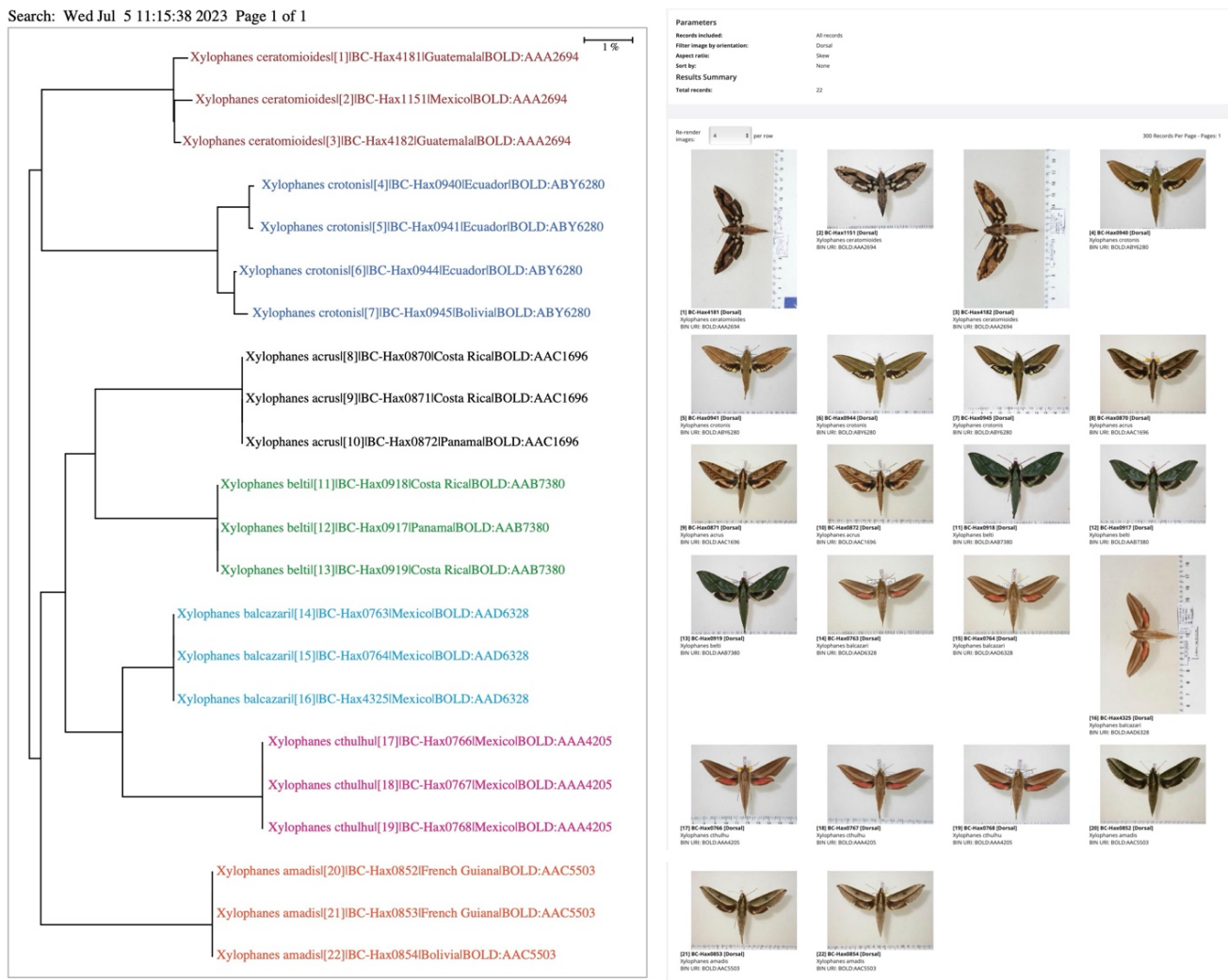

**Figure 9**. Tree Analysis with Matching Images. The left panel displays the Neighbor Joining tree, resulting from the Taxon ID analysis of 50 records belonging to the genus of *Xylophanes*(Lepidoptera: Sphingidae) (project code: JHPUB). The tree is colour-coded to represent the different BINs. The right panel shows the corresponding Image library produced using the Matching Images feature.

## 4.6 Dataset Assembly

BOLD supports the creation of customizable datasets that can, for example, assemble records from a particular taxonomic group or geographic region. Because a dataset can assemble both public and private (with permission) records, users have considerable flexibility in constructing datasets. Although these assembled datasets serve various functions, their primary utility lies in streamlining data publication.

The construction of a dataset for publication often necessitates the extraction of validated records from several projects. Because of variability in the reliability of records, it is best practice to only select records with high quality and accuracy for inclusion in a dataset.

To assemble a specific set of records, one must first establish a dataset within BOLD. This process parallels that for the creation of a project as it is executed from the Main Console. Each new dataset must be given both a unique dataset code as an identifier and a concise description.

Once a dataset has been established, users can add records to it. For private records, users must request access to each project with relevant data, navigate to the record list, and select the records of interest. Integrating the chosen records into the dataset is completed by simply selecting Add To Dataset option in the main menu on the left of the Record Console.

The expansion of a dataset with new records can occur progressively as they meet the requirements for inclusion. BOLD also allows for the inclusion of previously published records through a comprehensive search function, supporting queries based on taxonomy, geography, and other parameters. The dataset function plays a key role in supporting BOLD's efforts to promote data reuse.

After a dataset has been assembled, records should be comprehensively validated before publication. This stage is critical as it helps to uncover and rectify any inconsistencies that emerge when records from diverse sources are compiled. Thus, this final validation step ensures that dataset's integrity and high quality.

## 4.7 Dataset Validation

The assembly and publication of DNA barcode datasets on BOLD demand rigorous validation to ensure data accuracy and reliability before their release. Numerous challenges can arise, including poor quality data stemming from errors in DNA extraction, PCR amplification, sequencing, or contamination resulting in non-target sequences. Species misidentification is another common problem, often reflecting limited access to taxonomic expertise. Furthermore, a lack of multiple independent sequences from a species limits the evaluation of intraspecific variation, a critical matter for the discrimination of closely allied species. Incomplete annotations, such as missing collection dates, locations, or collector's names, can also impede data interpretation.

BOLD offers four tools to aid dataset validation before publication: (i) BIN Discordance, (ii) Barcode Gap Analysis, (iii) Project Summary, and (iv) Batch ID Engine.

The BIN Discordance tool (**Figure 10**) can only be applied to BINs with two or more records. It identifies discrepancies in taxonomic assignments among the records in a BIN, which may arise from misidentification or contamination during analysis.

Complementing the BIN Discordance tool is the Barcode Gap Analysis tool (**Figure 11**), which compares intraspecific and interspecific divergences for each species. As intraspecific divergences are usually much lower than interspecific ones[33], deviations warrant investigation as potential errors or biological explanations such as recently diverged species[34].

The Project Summary tool (**Figure 12**) checks record completeness. It generates a report detailing the distribution of collection locations and a species table, pinpointing gaps in coordinates, taxonomy, and sequences. This tool supports publications associated with reference libraries by providing supplementary materials.

The Batch ID Engine tool (**Figure 13**) assesses the accuracy of taxonomic assignments within a dataset. As it performs cross-validation with other sequences on BOLD, it is advisable to run all datasets through this tool before publication. Cases where queried records match different taxa often indicate analytical mishaps. Other mismatches may reflect errors in published datasets; such cases should be reported to the BOLD support team.

Compiling a DNA barcode library is a key task as it creates a reference for other researchers. Errors can lead to the replication of misidentifications and subsequent issues, emphasizing the importance of careful validation in creating a high-quality library.

**Figure 10.** BIN Discordance Report Summary: BIN discordance analysis for records of the order Hemiptera from Canada in the dataset DS-HECALIB. The orange rectangle highlights records that exhibit taxonomic discordance.

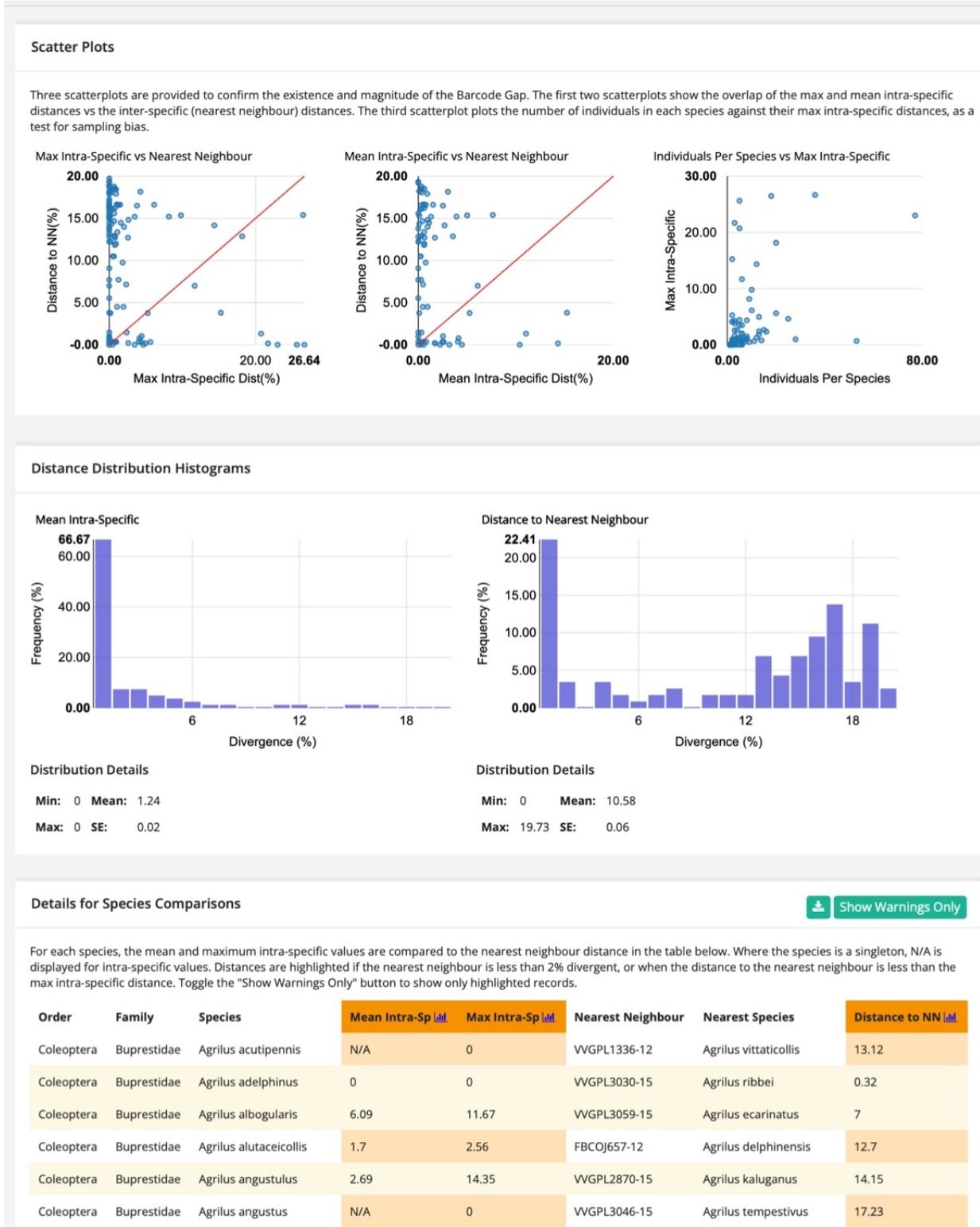

**Figure 11.** Barcode gap analysis.

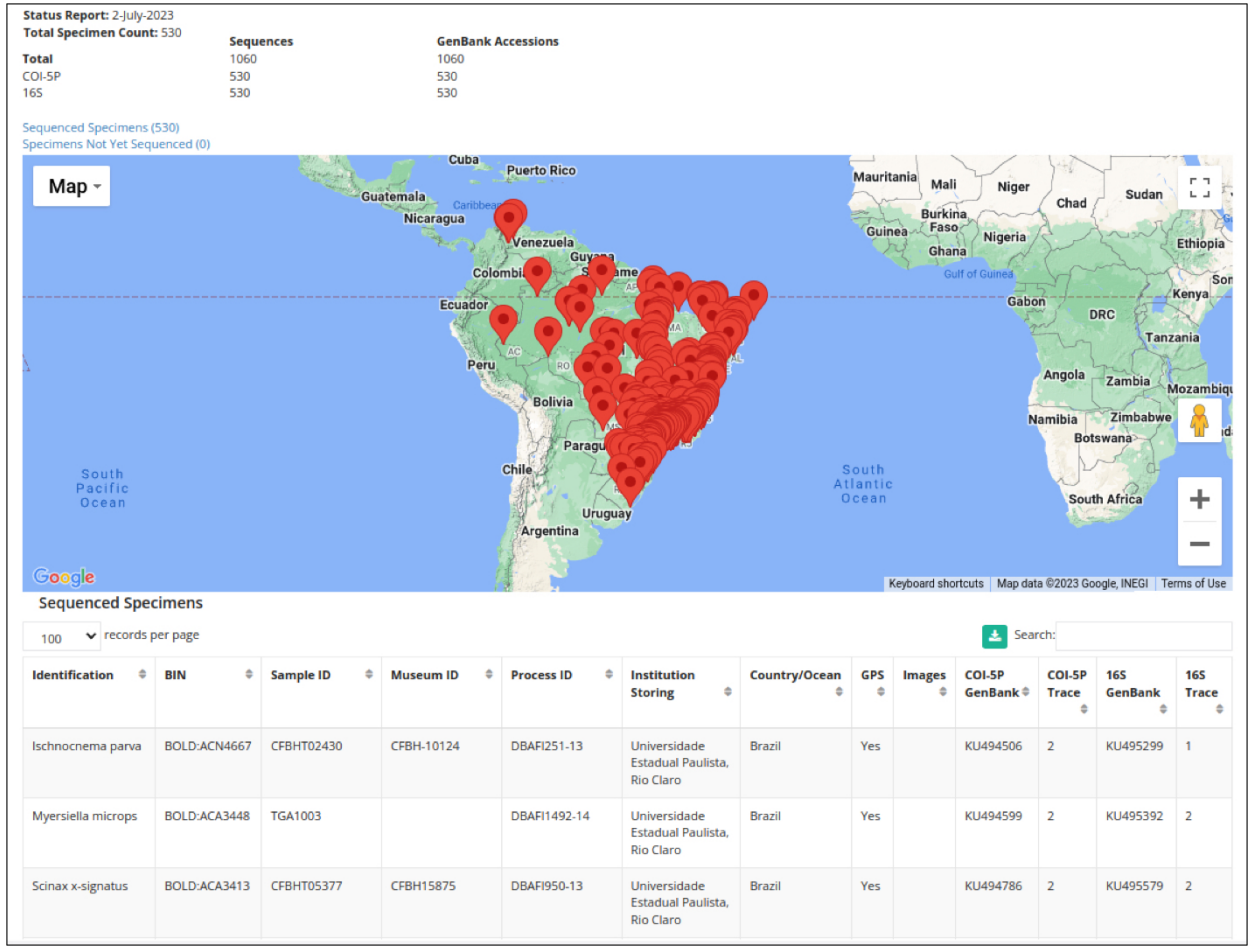

**Figure 12.** Project Summary.

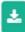

**Figure 13.** Results from the Batch ID Engine.

### 4.8 Dataset Publication

The publication of BOLD records as accessible datasets is the final step in the lifecycle and an important step in making data widely available. Offering a path to enhance data visibility, this process changes data access through the BOLD workbench by shifting the access level associated with the dataset to public status in the Dataset Options in the main menu. This raises accessibility by making the data available to all users of BOLD, whether for direct access or retrieval via searches.

This publication of datasets is only the first step in achieving data visibility. The subsequent recommended measure involves the association of each dataset with a publication via a DOI. Offering unique identification, convenient data sharing, and seamless access, the DOI also ensures appropriate citation in associated research efforts, consolidating recognition of the dataset's role in research. BOLD supports the minting of DOIs for each dataset at the time of publication.

Many scientific journals require the submission of sequence records to GenBank before publication.

BOLD eases compliance by integrating a data exchange pipeline with the National Center for Biotechnology Information (NCBI), facilitating data transformation into GenBank formats, and performing the submission.

Following submission, GenBank accessions are shared with the submitters for incorporation in publications. An embargo period is established on GenBank to accommodate the time for peer review.

The final step following acceptance by a journal is to associate the citation with the dataset on BOLD, to boost visibility of both the data and the publication. This association is accomplished via the publication interface in the main menu of the Dataset Console.

By 2022, BOLD had published 794 datasets with attributed DOIs, demonstrating its dedication to making data available to the scientific community and acknowledging the contributions of researchers. This successful cycle of record compilation, validation, and publication emphasizes the BOLD's integral role in the global DNA barcoding effort.

## 5 Integrated Analytics

BOLD has a unique configuration that enables direct analytical operations on data in it (**Table 9**). This incorporation of analytical functionalities affords several advantages. Chief among them is the ability for frictionless and immediate analysis. Data can be efficiently scrutinized right after it is added or updated within the system.

| Tool | Functionality |
|---|---|
| Taxon ID Tree | Interactive generation tool of phylogenetic hypothesis using the Neighbor Joining algorithm with various export options, including rooted or unrooted taxon ID trees, circular phylograms, and cladograms[35]. |
| Distance Summary | Reports divergence between sequences at the conspecific, congeneric, and confamilial levels. |
| Sequence Composition | Frequency of DNA bases, observed with emphasis on GC-content. |
| Barcode Gap Summary | Nearest Neighbor distances for each species in a list of specimens. |
| Diagnostic Characters | Polymorphisms between sets of nucleotide or amino acid sequences to characterize consensus characters by their diagnostic potential. |
| Cluster Sequences | Operational Taxonomic Unit (OTU) generation from identified or unidentified sequences using the Refined Single Linkage algorithm (RESL)[36], providing an initial estimate of species diversity and specimen affinities. |

| Geo-Distance Correlation | Correlation investigation using two methods: Mantel Test[37] between the geographic distance matrix (in kilometres) and the genetic divergence matrix. Spread comparison with minimum spanning tree of collection sites and maximum intraspecific divergence[22]. |
|---|---|
| Accumulation Curve | Diversity estimation and sampling efficiency of areas of collections using standardised DNA barcodes. |
| Diversity Measures | This tool provides insights into two key aspects of biodiversity: Alpha and Beta diversity. Alpha diversity focuses on the species richness in a specific set of data, helping to understand the diversity within a sample or at a location. Beta diversity examines the turnover or differentiation of species composition among multiple sites or samples. It assesses the degree of similarity or dissimilarity in species composition between different locations or samples[38]. |
| Batch ID Engine | Sequence matches based on similarity to entries in the BOLD database. |
| BIN Discordance Report | BIN records are shown by degree of conflict by comparing the taxonomy on selected records against others in the BIN records they are associated with. |

**Table 9.** The 11 tools on BOLD for analyzing specimen and sequence data. All can accessed from the project or dataset consoles.

Another advantage lies in the simplicity of data analysis. This approach avoids the need for data conversion steps during analysis, making the process seamless. BOLD's analytical tools are designed to align with its data model, ensuring a fluid analytical process. BOLD also provides a functionality that promotes the sharing and distribution of analytical results. This reflects two complementary mechanisms: its capacity to store results and access them via URL and the inherent communal nature of BOLD - where users may run analyses on data they can access.

While analysis is typically conducted on a project-by-project basis, BOLD allows the consolidation of records from multiple projects into "merged" projects, allowing more comprehensive examination. The system also supports data searches within BOLD or specific projects (**Figure 14**) based on a wide range of criteria, including lists of identifiers (Process IDs, Sample IDs, BINs), taxon names, geographic areas, tags, and collection dates. This search capability enables users to focus on specific subsets of the data to investigate localized patterns.

Finally, it is important to note that size restrictions apply to certain analytical elements, like the Neighbor Joining (NJ) tree which limits analysis to 25,000 records. However, other analytical tools, like the ID engine, scale linearly so they can handle all data records. These integrated analytical functions allows BOLD to aid the advance of DNA barcoding research.

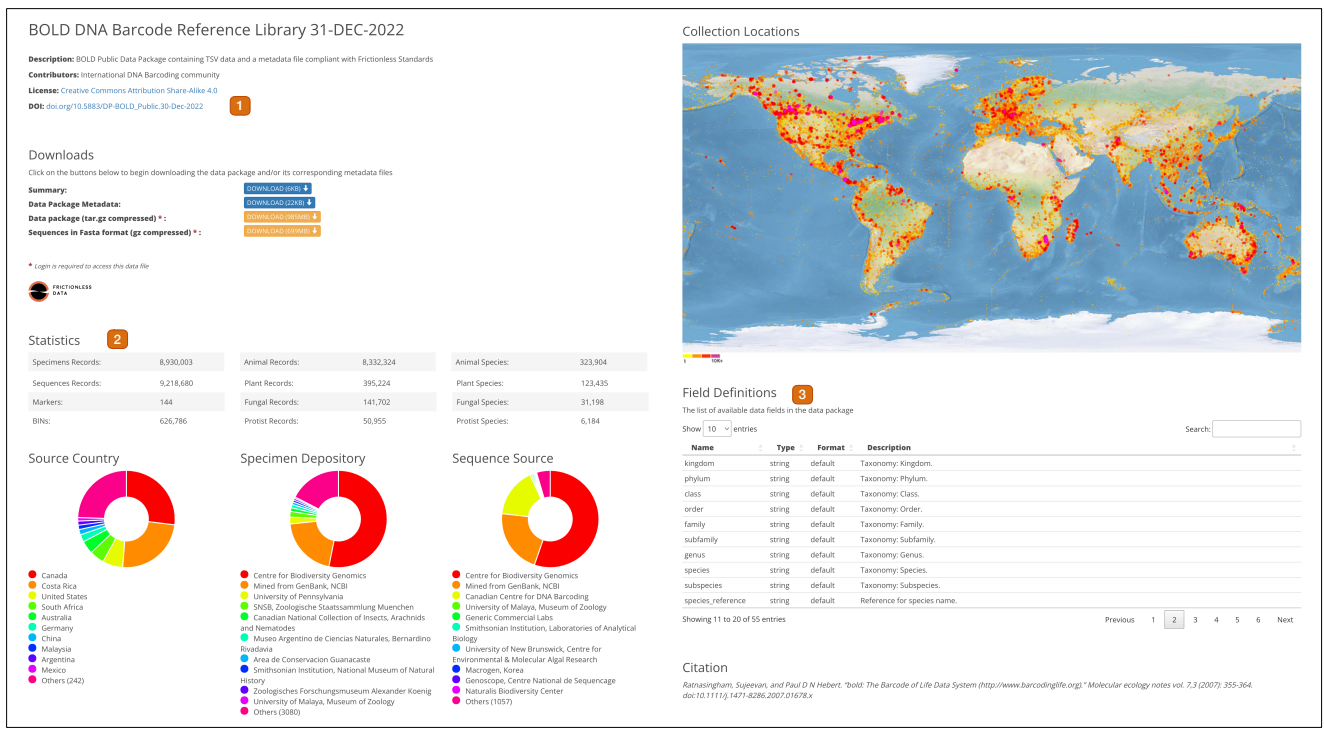

**Figure 16.** The data package for the BOLD library for December 31, 2022 shows the following key features: (1) Digital Object Identifier; (2) Summary statistics, encompassing the number of records, sequences, markers, and BINs; (3) Data dictionary.

## 6 Data Access

BOLD hosts public-facing search tools that allow users to query and extract data from four different databases. Two of these databases, the Public Data Portal and the BIN database, provide access to primary data on BOLD. The data in these databases is open to any user while private data is masked and only used for a statistical overview. These search tools and databases collectively allow researchers to access a wide range of sequence data and supplementary information.

### 6.1 Public Data Portal

The Public Data Portal is the central repository for all public data in BOLD. It grants users the capacity to explore and retrieve every public record in the database, making it a key resource for work on any taxonomic group. The portal returns all information associated with each record, providing users with a comprehensive view of the data. There are two ways to access the Public Data Portal: the Databases link and the Taxonomy Browser, when searching for a particular taxon. The Public Data Portal should be the first stop for users as they begin research on a specific taxon as it is the gateway to the reservoir of data on BOLD.

The Public Data Portal allows users to perform simple searches based on keywords, such as taxonomy, geography, specimen depository, project or dataset code, and specimen and sequence identifiers. More complex queries, involving the combination of two or more keywords, can also be constructed. For example, the following keywords Argentina[geo] Aves[tax] "Museo Argentino de Ciencias Naturales" (**Figure 15**) would return all public records in BOLD for Argentinian birds which are held in the Argentinian Museum of Natural Sciences. Users can also search for BINs directly. The data obtained through these searches can be downloaded in varied formats (TSV, DWC, XML, FASTA) along with related trace files. Additionally, maps can be generated from the data and downloaded at three resolutions. There is a limit of 100,000 records that can be downloaded at a time. **Figure 15** provides an overview of the Data Portal and its capabilities.

## 6.2 Data Packages

The rapid growth of the BOLD database and the rising adoption of DNA barcoding have led to a strong demand for immediate access to large datasets. For many users, accessing data through portals or Application Programming Interfaces (APIs) is inefficient for large-scale requirements. Bulk downloading, particularly within institutions with multiple users, is more effective, ensuring consistent access to datasets.

Additionally, methods like metabarcoding and environmental DNA (eDNA) analyses require copies of comprehensive datasets for accurate data analysis. These analyses need access to extensive and reliable datasets, whether run on local systems or cloud-based platforms.

To address these needs, the Data Packages module was introduced. This module serves packaged datasets in as single compressed files with supporting metadata, increasing the discoverability and accessibility of data for users. A key feature of the Data Packages module is its adherence to the Frictionless Data standard[39] created by Open Knowledge Foundation. This standard offers a structured method for data description, packaging, and validation, using a container format to combine structured data with its associated metadata in JSON format. This format was chosen to simplify data sharing, validation, and interoperability.

The approach to package assembly is systematic. All public records are compiled into snapshots on a quarterly basis, with each package summarized on a dedicated page (**Figure 16**) and referenced by a DOI. For users needing the most up-to-date data, BOLD produces weekly snapshots. Project and program-focused data packages are also supported. One significant package in this category is from the iBOL's BARCODE 500K program.

## 6.3 Taxonomy Browser

The BOLD Taxonomy Browser (**Figure 17**) is indispensable for researchers engaged in constructing DNA barcode libraries. It provides a summary of barcode coverage for specific taxa and facilitates exploration of the taxonomic hierarchy from the kingdom down to the species. Accommodating entries for animals, plants, fungi, and protists, the browser is a valuable tool for diverse research areas.

Because the Taxonomy Browser is integrated with the BOLD and Barcode Index Number (BIN) databases, users can access associated specimen and sequence records for any taxon. In this way, the browser acts as a dynamic reflection of BOLD's current state, encapsulating the latest updates in taxonomy.

The browser allows access to varied data elements from summary statistics like the number of records and species for a particular taxonomic group to more detailed information such as associated BINs and individual records. Additionally, it provides information about the distribution of collection sites and the locations of specimens, enhancing the contextual understanding of the data.

The Taxonomy Browser also displays representative images for each taxon, enabling the visual verification of current assignments. Its most significant role lies in reducing the duplication of effort in library construction, enabling researchers to monitor progress on taxa of interest.

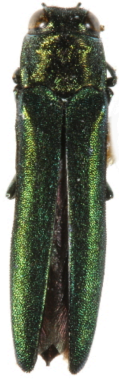
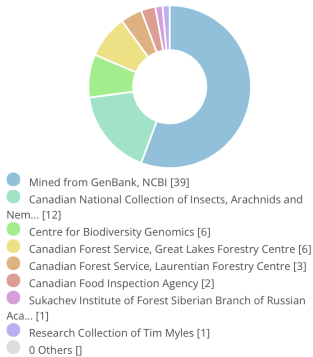
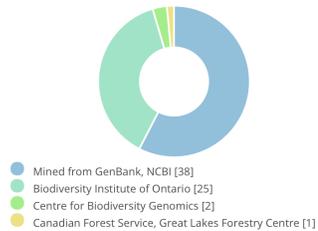
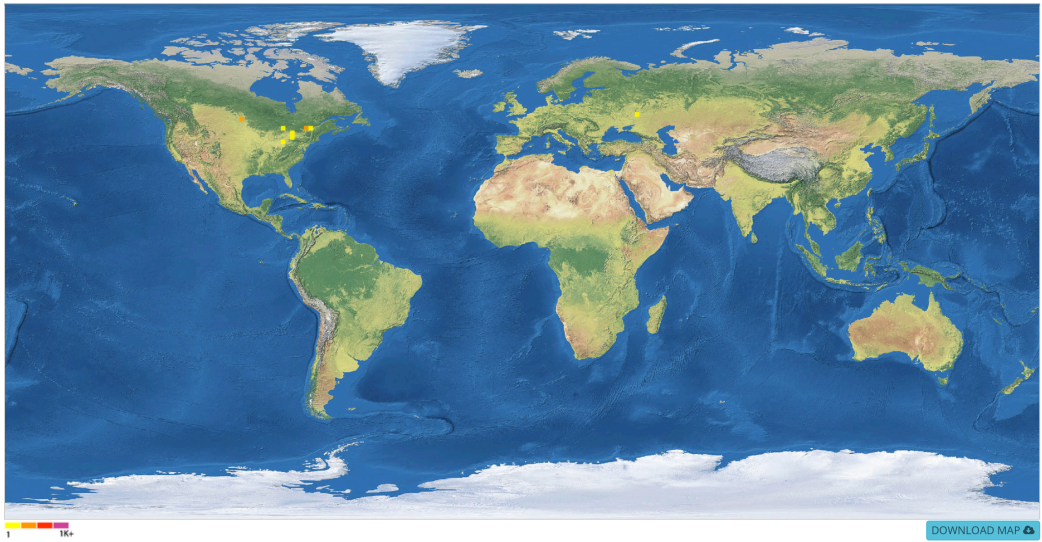
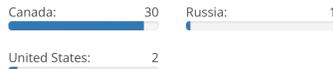
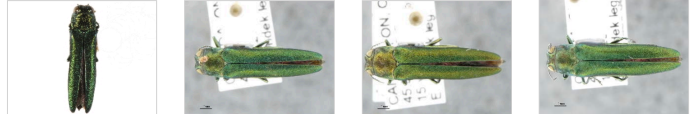

**Figure 17.** The Taxonomy Browser page for *Agrilus planipennis* (Emerald Ash Borer Beetle) displays: (1) Coverage statistics; (2) Specimen repositories; (3) Laboratories responsible for sequence generation; and (4) Collection sites.

## 6.4 Application Programming Interface (APIs)

The BOLD API streamlines access to the DNA barcode sequences, specimen data, and associated metadata. Via the API endpoints (**Table 10**), users can programmatically retrieve data subsets based on various criteria, including taxonomy, geography, or specific gene markers, aiding more comprehensive, automated analyses. Supporting various data formats (i.e., XML and JSON), it ensures broad compatibility and easy integration with other bioinformatics tools/platforms. Notably, the API has been utilized by ROpenSci[40], an initiative that promotes open and reproducible research using shared data and reusable software. Their integration of the BOLD API facilitates easier access and analysis of barcoding data within the R environment, a widely used programming language and software environment for statistical computing. The API also offers an analytical endpoint that enables near real-time DNA barcode-based species identifications, bridging the gap between raw sequence data and practical applications. Developers can leverage this API to build custom applications, tools, or plugins, to enhance the utility of BOLD data in diverse contexts.

| API endpoints | Description |
| --- | --- |
| Summary Statistics | Retrieves summary information provided by BOLD public searches.<br><br>e.g., *http://www.boldsystems.org/index.php/API_Public/stats?taxon=Eacles&geo=Mexico&format=tsv* |
| Specimen Data | Retrieves data records matching any combination of the following 7 parameters: taxon, IDs, BIN, container, institution, researcher, geolocation.<br><br>e.g., *http://www.boldsystems.org/index.php/API_Public/specimen?taxon=Aves&geo=Costa%20Rica&format=tsv* |
| Sequence Data | Retrieves sequences matching any combination of the following 9 parameters: taxon, IDs, BIN, container, institution, researcher, geolocation, gene marker, format.<br><br>e.g., *http://www.boldsystems.org/index.php/API_Public/sequence?taxon=Chordata&geo=Florida&institutions=Smithsonian%20Institution* |
| Specimen and Sequence data | Retrieves specimen data and sequence records matching any combination of the following 9 parameters: taxon, IDs, BIN, container, institution, researcher, geolocation, gene marker, format.<br><br>e.g., *http://www.boldsystems.org/index.php/API_Public/combined?taxon=Mammalia&geo=Canada&format=tsv* |

| Sanger Trace Files | Retrieves trace files matching any combination of the following 9 parameters: taxon, IDs, BIN, container, institutions, researcher, geolocation, gene marker, format.<br><br>e.g., *http://www.boldsystems.org/index.php/API_Public/trace?taxon=Bombus&institutions=York%20University* |
|---|---|
| Taxonomy ID Service | Retrieves taxonomic information generated by BOLD taxonomy ID.<br><br>*http://www.boldsystems.org/index.php/API_Tax/TaxonData?taxId=88899&dataTypes=basic,stats* |
| Taxon Name Service | Retrieves taxonomy information by taxon name.<br><br>e.g., *http://www.boldsystems.org/index.php/API_Tax/TaxonSearch?taxName=Diplura* |
| Rest API for ID Engine | The BOLD ID Engine Web Service can query the COI ID Engine via URL. An XML file with the top public matches (up to 100) is retrieved by querying a COI sequence.<br><br>e.g., *http://www.boldsystems.org/index.php/Ids_xml?db=COX1_SPECIES_PUBLIC&sequence=TATAAATAATATAAGATTTTGACTTCTTCCACCTTCTTTAACTCTTCTCCTATCCAGAGGAATAGTTGAAAGAGGTGTTGGCACAGGATGAACTGTTTATCCTCCTTTAGCTGCTGGAATCGCCCATGCAGGCGCTTCTGTGGACTTAGGAATTTTTTCTCTTCATATAGCGGGAGCTTCTTCTATTTTAGGGGCGGTAAATTTTATTACTACT* |

**Table 10.** List of the nine endpoints/functions of the BOLD API.

## 7 Conclusion

The BOLD platform has been a pivotal component of biodiversity research since its inception. As an essential part of the global initiative for species identification using DNA barcoding, BOLD has consistently played a vital role in consolidating and managing molecular and specimen-based data. Providing an interface for collecting, storing, analyzing, and publishing DNA barcode records, effectively bridging the gap between genetic data and taxonomic knowledge. Over nearly two decades of operation, the BOLD system has expanded in sophistication, scope, and utility, becoming an integral component of numerous research initiatives and programs.

**Acknowledgement**

We are profoundly grateful to the funders which have supported BOLD since its inception in 2004. They include the Canada Foundation for Innovation, Genome Canada, Gordon and Betty Moore Foundation, Natural Sciences and Engineering Research Council of Canada (NSERC), Ontario Ministry of Research and Innovation, Ontario Genomics, Canada First Research Excellence Fund, and the New Frontiers in Research Fund.